\documentclass[twocolumn,english,prl,epsfig,rotate,showpacs,aps, superscriptaddress]{revtex4-2}
\usepackage[T1]{fontenc}
\usepackage[latin9]{inputenc}
\usepackage{color}
\usepackage{epsfig}
\usepackage{verbatim}
\usepackage{amsmath}
\usepackage{amsthm}
\usepackage{amssymb}
\usepackage{graphicx}
\usepackage{mathrsfs}

\usepackage{cancel}
\usepackage{esint}
\usepackage{microtype}
\usepackage{dcolumn}
\usepackage{bm}
\usepackage{amsfonts}
\usepackage[table]{xcolor} %
\newcommand{\blue}{\color{black}}

\newcommand{\bluer}{\color{black}}
\makeatletter
\usepackage{babel}
\usepackage{braket}

\makeatother

\usepackage{babel}

\begin{document}

\title{Layer-number parity induced topological phase transition}

\author{Kai Chen$^{\chi}$}
\affiliation{School of Physics Science and Engineering, Tongji University, 200092 Shanghai, China}

\author{Junyan Guan}
\affiliation{School of Physics Science and Engineering, Tongji University, 200092 Shanghai, China}

\author{Jiamin Guo}
\affiliation{School of Physics Science and Engineering, Tongji University, 200092 Shanghai, China}

\author{He Gao}
\affiliation{School of Advanced Manufacturing Engineering, Nanjing University, 215163 Suzhou, China}

\author{Zhongming Gu$^{*}$}
\affiliation{School of Physics Science and Engineering, Tongji University, 200092 Shanghai, China}

\author{Jie Zhu$^{\dagger}$}
\affiliation{School of Physics Science and Engineering, Tongji University, 200092 Shanghai, China}

\date{\today}

\begin{abstract}
We demonstrate that stacking topologically trivial layers, under enforced symmetry restrictions, yields emergent topological phases with protected boundary states. Remarkably, the number of layers itself acts as a topological switch, enabling the system to host topological bound states in the continuum (BICs).  {\blue{Analytically, we reveal that an odd-layer configuration not only renders the spectrum gapless but also supports BICs protected by interlayer reflection symmetry}}. Combined with entanglement-spectrum calculations, this confirms that odd-layer systems indeed support topological BICs. We provide experimental confirmation of these topological states in stacked acoustic lattices. Our findings establish a previously overlooked pathway to topology and demonstrate a readily applicable strategy for realizing exotic states in a wide range of artificial material systems.
\end{abstract}
\maketitle

\textit{Introduction}---Topological phases of matter are characterized by properties that remain invariant under continuous deformations, distinguishing them from trivial phases through quantized invariants protected by symmetry or by the closing of a bulk energy gap \cite{hasan2010colloquium,qi2011topological}. These phases host robust boundary excitations, such as the chiral edge channels of the quantum Hall effect \cite{von202040} and the helical edge modes of quantum spin Hall insulators \cite{bernevig2006quantum,maciejko2011quantum}. The celebrated tenfold way classification, based on time-reversal, particle-hole, and chiral symmetries, provides a systematic framework for categorizing such topological phases across dimensions \cite{kitaev2009periodic,ryu2010topological,chiu2016classification}. Beyond gapped systems, gapless topological phases---including Dirac and Weyl semimetals---feature symmetry-protected degenerate points that give rise to exotic surface states and electromagnetic responses \cite{hosur2013recent,yan2017topological,armitage2018weyl}. In interacting systems, topologically ordered phases host fractionalized excitations, such as the anyonic quasiparticles of the fractional quantum Hall effect, which are promising for fault-tolerant quantum computation \cite{stormer1999nobel,moore1991nonabelions}.

In recent years, the principles of topological physics have been successfully extended to engineered classical wave systems---including photonic \cite{ozawa2019topological,lu2014topological,khanikaev2017two}, acoustic \cite{yang2015topological,xue2022topological}, and mechanical metamaterials \cite{bertoldi2017flexible,xin2020topological}---offering a highly tunable platform for investigating and harnessing topological phenomena. The interplay of topology and symmetry provides a powerful framework for designing functional devices that leverage topologically protected states, such as wave propagation along edges that is robust against disorder and defects \cite{wu2015scheme,bliokh2015quantum,ding2019experimental,yang2019realization,rechtsman2013photonic,price2022roadmap,alu2025bright}. The flexibility of classical systems has enabled the exploration of higher-order topological phases \cite{benalcazar2017quantized,benalcazar2017electric,benalcazar2019quantization,schindler2018higher,chen2021nonlocal}, where boundary states localize at corners or hinges rather than edges \cite{ezawa2018higher,ni2019observation,xue2019acoustic,kim2020recent,li2020higher,yang2020helical,mandal2025photonic}. These phases enable novel device functionalities, including topological lasers with single-mode operation \cite{kim2020multipolar,wu2023higher,rao2024single} and low threshold \cite{zhang2020low}, as well as acoustic sensors, energy harvesters, and focusing devices \cite{zhang2019second,xie2021higher}.
\begin{figure}[h]
\includegraphics[width=0.9\columnwidth,height=0.7\textheight,keepaspectratio]{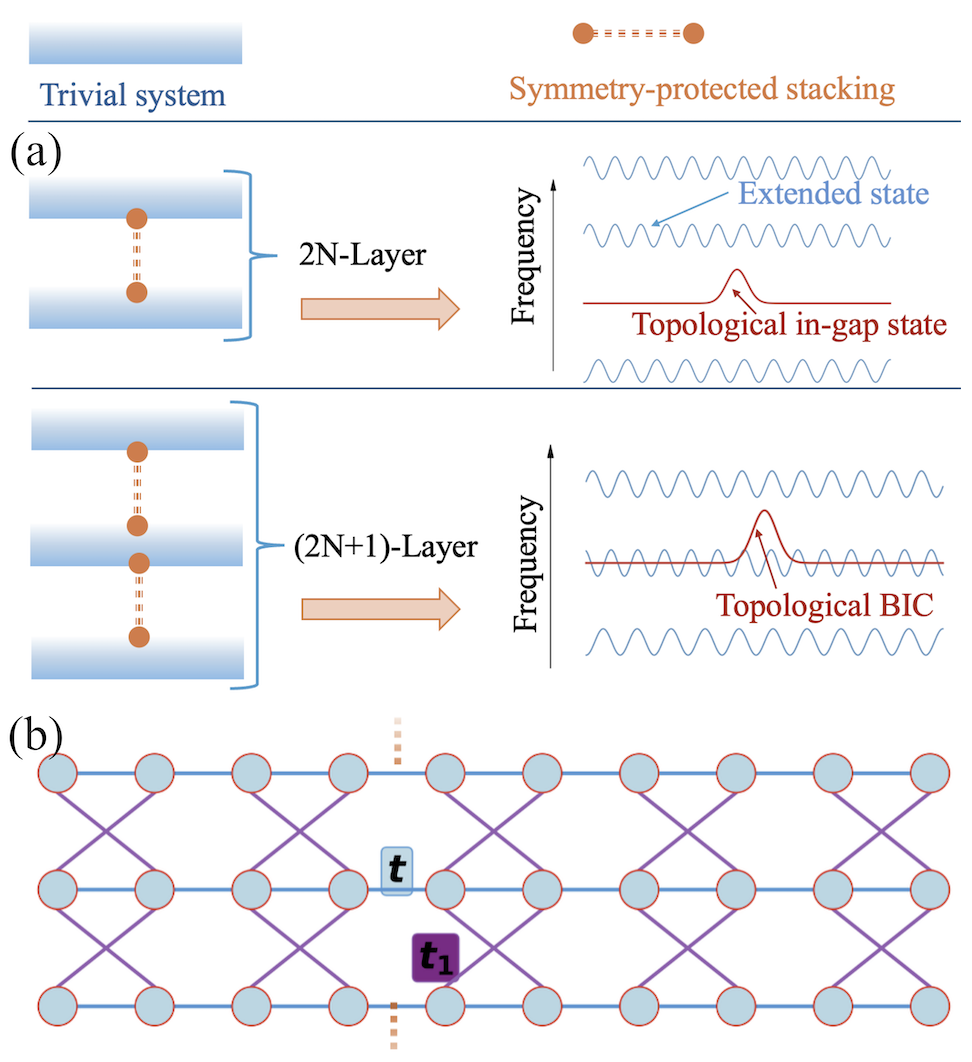}
\caption{(Color online) Evolution of topological states via layer stacking. (a) Schematic illustrating trivial-to-topological phase transitions. Symmetry-constrained stacking in $2N$-layer and $(2N+1)$-layer systems drives the evolution of topological states from isolated in-gap edge modes to BICs embedded within the bulk continuum. (b) Lattice structures of the stacked systems, where $t$ and $t_1$ denote the intralayer and interlayer coupling amplitudes, respectively.}
\label{fig0}  
\end{figure}
A particularly intriguing development in this field is the emergence of topological bound states in the continuum (BICs): localized states whose eigenvalues lie within the continuous spectrum of extended modes \cite{zhen2014topological,hsu2016bound,cerjan2020observation,liu2023universal,hu2021nonlinear,bulgakov2017topological,qian2024non}. These states combine topological protection with theoretically infinite quality factors, offering unprecedented opportunities for wave trapping, sensing, and lasing \cite{yin2025dimensional,guo2024realization,deriy2022bound,kang2023applications}. However, most realizations of topological BICs have relied on complex geometries or delicate parameter tuning. While topological BICs can be realized by mirror-stacking existing topological phases \cite{liu2023universal}, a fundamental question remains: is it possible to induce topological BICs by stacking systems that are individually trivial?

In this letter, we introduce a symmetry-engineering approach to systematically induce topology---including topological BICs---in stacked trivial systems. While moir\'{e} engineering via twisting has emerged as a powerful tool for modulating electronic structures in van der Waals materials \cite{bistritzer2011moire,cao2018unconventional,andrei2021marvels}, symmetry manipulation often provides a more direct and versatile pathway in noninteracting classical platforms. By designing interlayer couplings that impose a global chiral symmetry---a property fundamentally absent in the individual constituent layers---and by varying the stacking layer number $N$, we access a rich landscape of topological phases. This approach yields both gapped phases characterized by protected in-gap edge states and, more notably, gapless phases hosting topological BICs {\blue{protected by the interlayer reflection symmetry}}, whose eigenvalues remain embedded within the bulk continuum. We validate these findings through a combination of numerical calculations, symmetry considerations, and entanglement spectrum (ES) analysis, culminating in the experimental observation of these states in 3D-printed acoustic lattices.

\begin{figure}[h]
\includegraphics[width=1\columnwidth]{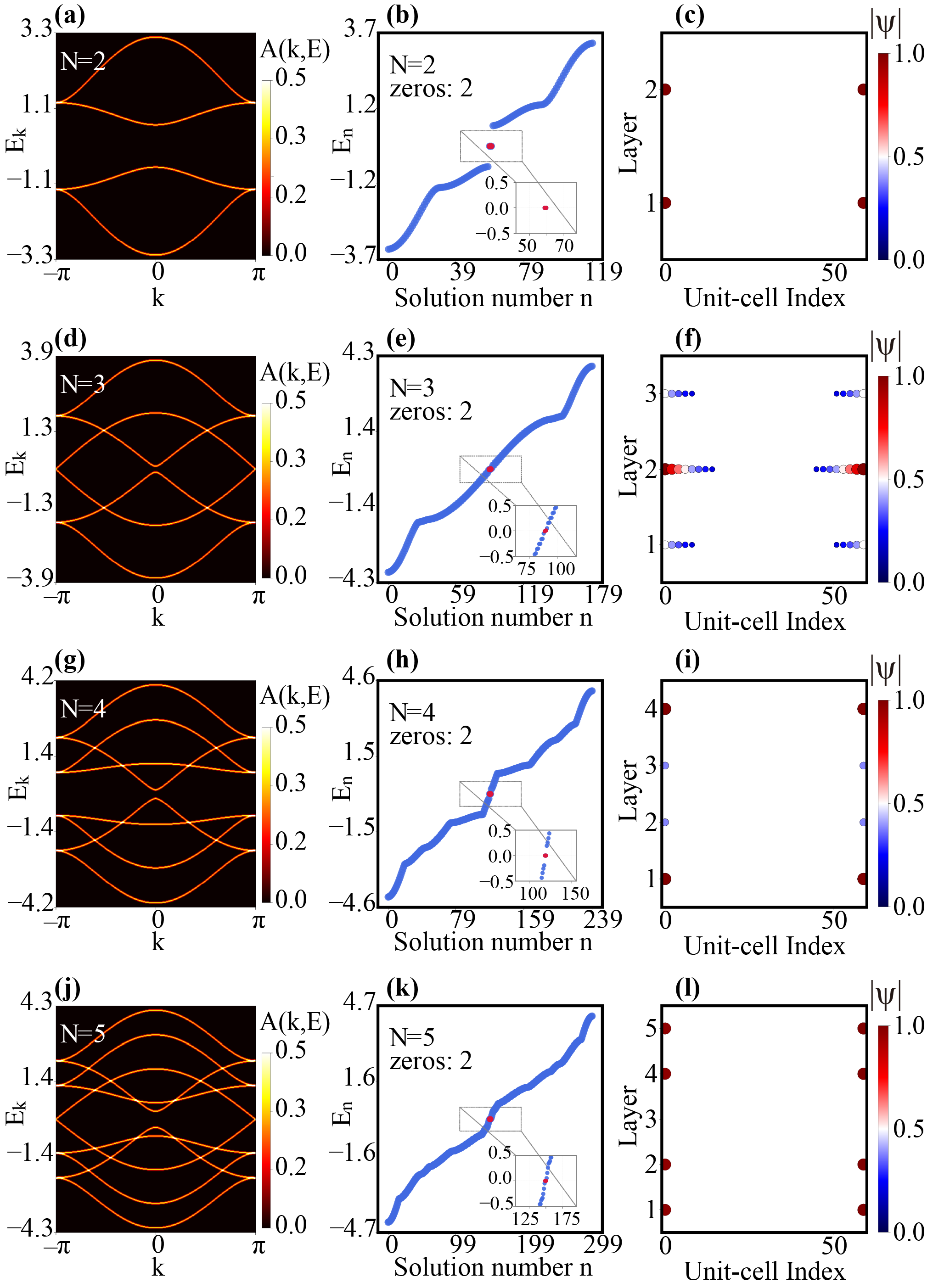}
\caption{(Color online) Transition between topological in-gap states and BICs in multi-layer systems.
(a--c) Bulk band structures weighted by the spectral function $\text{A(k, E)}$, energy spectra for finite systems under open boundary conditions, and spatial distribution of the magnitude of characteristic wavefunctions $|\psi|$ for 2-layer configurations. 
(d--f), (g--i), and (j--l) show corresponding results for 3, 4, and 5-layer configurations, respectively. 
The topological states (red dots) appear as in-gap edge states for even-layer systems ($N=2, 4$) and as BICs embedded within the continuum for odd-layer systems ($N=3, 5$). 
Parameters: $t = 1$, $t_1 = 1.34$.}
\label{fig1}
\end{figure}

\textit{Topological Modes in Stacked Trivial Systems}---Previous studies have shown that stacking topological systems can lead to new topological phases, with intriguing examples including the realization of second-order topological phases with corner states by stacking topological SSH chains \cite{wei2023topological,guo2024realization,bongiovanni2024p,luo2023higher,duan2025tunable,luo2025family}. However, stacking \textit{trivial} systems typically yields a trivial phase. Nontrivial topology can nonetheless be induced through engineered interlayer coupling or modified stacking geometry. In this work, we employ a symmetry-engineering approach that imposes chiral symmetry via tailored interlayer couplings. Starting from a one-dimensional metallic chain---a trivial building block with Bloch Hamiltonian \(H(k)=2t\cos k\)---we stack \(N\) copies and introduce coupling between them to form an \(N\)-layer system described by the Hamiltonian

\begin{equation}
H_N(k) = I_N \otimes H_{\mathrm{onsite}} + \bigl( I_{\triangle} \otimes h_t + \mathrm{h.c.} \bigr),
\label{eq:HN}
\end{equation}
where \(I_N\) is the \(N \times N\) identity matrix and \(H_{\mathrm{onsite}} = t(1+\cos k)\sigma_x + t \sin k\,\sigma_y\), with \(\sigma_{x,y,z}\) denoting the Pauli matrices. Here, \(I_{\triangle}\) is an \(N \times N\) strictly upper triangular matrix whose elements are given by \(I_{\triangle}^{mn}=\delta_{m+1,n}\). The term \(h_t = t_1 \sigma_x\) describes the specific interlayer coupling, where \(t_1\) sets the coupling strength. 

{\blue{The Hamiltonian in Eq.~\eqref{eq:HN} satisfies chiral symmetry, $\Gamma H_N(k) \Gamma^{-1} = -H_N(k)$, with $\Gamma = I_N \otimes \sigma_z$. In addition to chiral symmetry, the model supports the interlayer reflection symmetry $R H_N(k) R^{-1} = H_N(k)$, where $R$ maps the $n$-th layer to the $(N-n+1)$-th layer. Crucially, the chiral symmetry ensures that systems with an odd number of layers are gapless (see Section~II of the Supplemental Material~\cite{SM} for a proof). Parallel to this, the interlayer reflection symmetry acts as the protection mechanism for the BICs within these odd-layer systems. 

Specifically, the cross-coupled layers can be mathematically decomposed into independent 1D chains with distinct interlayer parities (see Sec.~IV of the Supplemental Material). {\bluer{To provide a more transparent theoretical interpretation, the mechanism should be viewed not merely as the stacking of trivial layers, but as the formation of effective SSH channels induced by cross-layer coupling. If the transverse direction is treated periodically, the two-dimensional Bloch Hamiltonian $H(k_x, k_y)$ yields an effective SSH intracell coupling of $v(k_y) = t + 2t_1 \cos k_y$, where the bulk gap closes continuously with $k_y$. For a finite stack of $N$ layers, these transverse modes are discretized into eigenchannels $\lambda_m = 2 \cos[m\pi/(N+1)]$, mapping the system into decoupled effective SSH channels with couplings $t_{\text{eff}} = t + \lambda_m t_1$. The role of the layer-number parity dictates the resulting physical classification. In the odd-layer case, a gapless bulk channel naturally appears through the exact $\lambda_m = 0$ transverse mode, while the topological boundary state originates from a separate transverse channel driven into the topological SSH regime ($|t_{\text{eff}}| < |t|$). Because these two channels inherit orthogonal spatial parities under interlayer reflection, their interaction matrix element strictly vanishes. Consequently, this state can be classified as a symmetry-protected topological BIC: the topological boundary state remains dynamically decoupled from the gapless transverse continuum strictly due to this channel orthogonality.}}

To further validate the existence of these BICs, we utilize non-equilibrium Green's function calculations~\cite{datta1997electronic} to demonstrate their dynamic decoupling. We find that while a symmetry-breaking single-layer excitation actively couples to the BIC to trigger a Fano resonance, a symmetry-preserving coherent excitation remains mathematically orthogonal to the state. Such an excitation bypasses the BIC entirely to yield perfect unit transmission (see Sec.~V of the Supplemental Material for details). This behavior, combined with the entanglement spectrum presented in the next section, further demonstrates that odd-layer systems host robust topological BICs.}}

In real space, the lattice Hamiltonian (depicted schematically in Fig.~\ref{fig0} (b)) is expressed as:
\begin{equation}
H_N = \sum_{i=1}^{L-1} \left(\Psi_{i}^\dagger T_{i,i+1} \Psi_{i+1} + \text{h.c.}\right) + \sum_{i=1}^{L} \Psi_{i}^\dagger T_{0} \Psi_{i},
\label{eq:HNR}
\end{equation}
where $\Psi_{i}^\dagger \equiv \left(\Psi_{i,1}^\dagger, \Psi_{i,2}^\dagger, \dots, \Psi_{i,N}^\dagger\right)$ is the creation operator vector in the $i$th unit cell, and $T_0$ and $T_{i,i+1}$ are the intra-cell and inter-cell coupling matrices, respectively. These coupling matrices are given by

\begin{equation}
\left\{
\begin{aligned}
T_0 &= t \, I_N \otimes \sigma_x + t_1 \left(I_{\triangle} \otimes \sigma_x + I_{\triangle}^\dagger \otimes \sigma_x\right), \\
T_{i,i+1} &= \frac{t}{2} \, I_{\triangle} \otimes (\sigma_x - i\sigma_y).
\end{aligned}
\right.
\label{eq:coupling}
\end{equation}

According to the tenfold way topological classification, a one-dimensional system with chiral symmetry belongs to class AIII and can host topological phases characterized by a winding number \cite{ryu2010topological}. After imposing the symmetry-constrained coupling, the number of layers \(N\) serves as a discrete tuning parameter. To explore the resulting topological states, we analyze the system under both periodic and open boundary conditions.

As shown in Fig.~\ref{fig1}, the bulk spectrum is gapless for $N=3$ and $5$ [Figs.~\ref{fig1}(d) and \ref{fig1}(j)], but gapped for $N=2$ and $4$ [Figs.~\ref{fig1}(a) and \ref{fig1}(g)]. Under open boundary conditions, in-gap topological edge states appear for the gapped cases ($N=2$ and $4$), as shown in Figs.~\ref{fig1}(b) and \ref{fig1}(h), with their corresponding localized profiles presented in Figs.~\ref{fig1}(c) and \ref{fig1}(i). This behavior aligns with the celebrated bulk-boundary correspondence, where robust topological edge states are determined by the bulk topology \cite{hasan2010colloquium,qi2011topological}. The dependence of these topological states on layer parity is further corroborated by COMSOL simulations, as detailed in Section III of the Supplemental Material~\cite{SM}.

Notably, the gapless systems ($N=3, 5$) also support edge states [Figs.~\ref{fig1}(f), (l)] that coexist with the bulk continuum [Figs.~\ref{fig1}(e), (k)], thereby constituting topological BICs. Unlike previous studies that generated topological BICs by embedding the boundary states of one topological subsystem into the bulk spectrum of another \cite{liu2023universal}, our approach realizes these states by stacking inherently trivial systems. This demonstrates a distinct mechanism for inducing bound states within a continuum through symmetry-protected layer stacking. 

\begin{figure}[h]
\includegraphics[width=1\columnwidth]{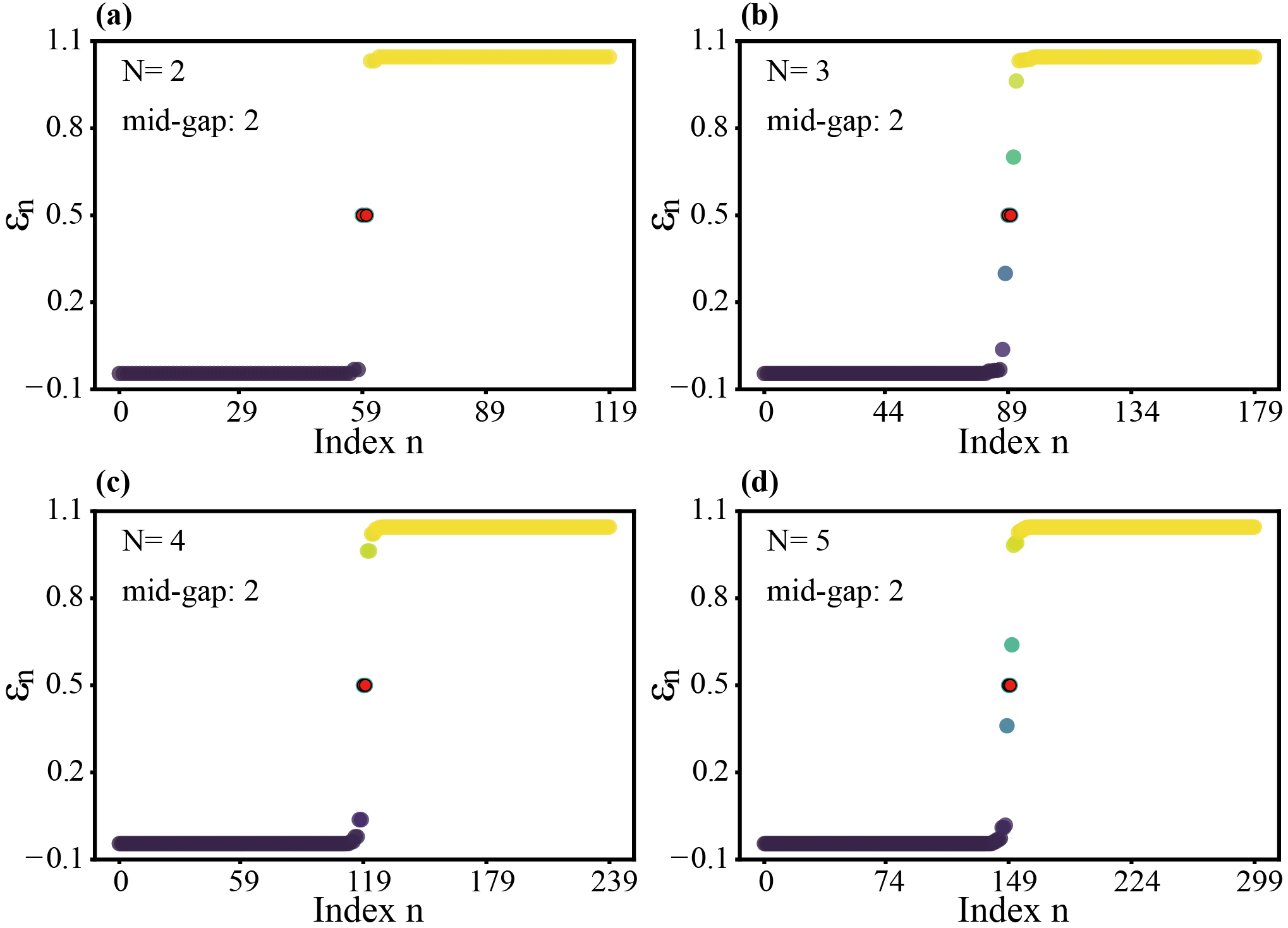}
\caption{(Color online) Entanglement spectrum $\varepsilon_n$ for 2-, 3-, 4-, and 5-layer systems, color-coded by value. All parameters match those in Fig.~2. The integers indicate the count of mid-gap states at $\varepsilon_n = 0.5$ (highlighted by red points). The total system consists of $L=60$ unit cells, with a subsystem size $L_A = L/2$.}
\label{fig2}  
\end{figure}

\textit{Topological entanglement spectrum}\textbf{---} The topology of a system can be characterized not only by bulk invariants (e.g., the Chern number \cite{thouless1982quantized,hatsugai1993chern}) but also by its ES \cite{li2008entanglement,peschel2009reduced,pollmann2010entanglement,turner2010entanglement,ryu2006entanglement,chang2020entanglement}. Beyond its use in detecting topological order, the ES and related entanglement entropy encode universal, conformally invariant information about critical systems and emergent spacetime structure \cite{calabrese2004entanglement,ryu2006holographic}. For our purposes, the ES is especially valuable for diagnosing gapless topological phases, where conventional bulk invariants may become ill-defined. In such cases, robust mid-gap states within the ES serve as unambiguous topological markers.

In a non-interacting system, the reduced density matrix for a subsystem $A$ takes a Gaussian form \cite{peschel2009reduced,chang2020entanglement}:
\[
\rho_A \propto \exp\left(-\sum_{\alpha\beta} \phi_{\alpha}^\dagger H_{\alpha\beta}^{E} \phi_\beta\right),
\]
where $H^{E}$ denotes the entanglement Hamiltonian, and $\phi_\alpha$ ($\phi_{\alpha}^\dagger$) are the annihilation (creation) operators for the single-particle state $\alpha$. The ES corresponds to the eigenvalues of the correlation matrix $C_{\alpha\beta} \equiv \operatorname{Tr}(\rho_A \phi_{\alpha}^\dagger \phi_{\beta})$ (see Section I of the Supplemental Material~\cite{SM} for details).

We now apply this formalism to the stacked system described in the previous section. The subsystem is chosen as half the total system size ($L_A = L/2$) by {\blue{cutting across the inter-cell bond}}, and the resulting entanglement spectra are presented in Fig.~\ref{fig2}. For the 2- and 4-layer systems in their topological phases---which host in-gap edge states under open boundary conditions---the entanglement spectra feature distinct mid-gap states [Figs.~\ref{fig2}(a) and (c)]. Their numbers coincide with the corresponding edge-state counts. Similarly, the entanglement spectra of the 3- and 5-layer systems---which host BICs under open boundary conditions---also exhibit mid-gap states, as shown in Figs.~\ref{fig2}(b) and (d), with their number matching the number of BICs. Furthermore, the presence of states between the mid-gap levels and the maximum or minimum of the spectrum signifies the gapless nature of the system and results in the logarithmic scaling dependence of the entanglement entropy \cite{chang2020entanglement}.

The entanglement spectrum thus provides a direct topological fingerprint for both gapped and gapless phases. In our stacked systems, the mid-gap state counts match exactly the numbers of protected boundary states---edge states for $N=2,4$ and BICs for $N=3, 5$---unambiguously establishing their topological nature even when conventional bulk invariants are ill-defined.

The study of topological phases in classical wave systems---kick-started by the robustness and unidirectional, backscattering-immune propagation of topologically protected states in platforms such as photonics, acoustics, and electrical circuits---has reshaped our understanding of wave propagation and enabled robust strategies for guiding, routing, and localizing waves. In the following section, we explore topological in-gap states and BICs in acoustic lattices.

\begin{figure}[h]
\includegraphics[width=1\columnwidth]{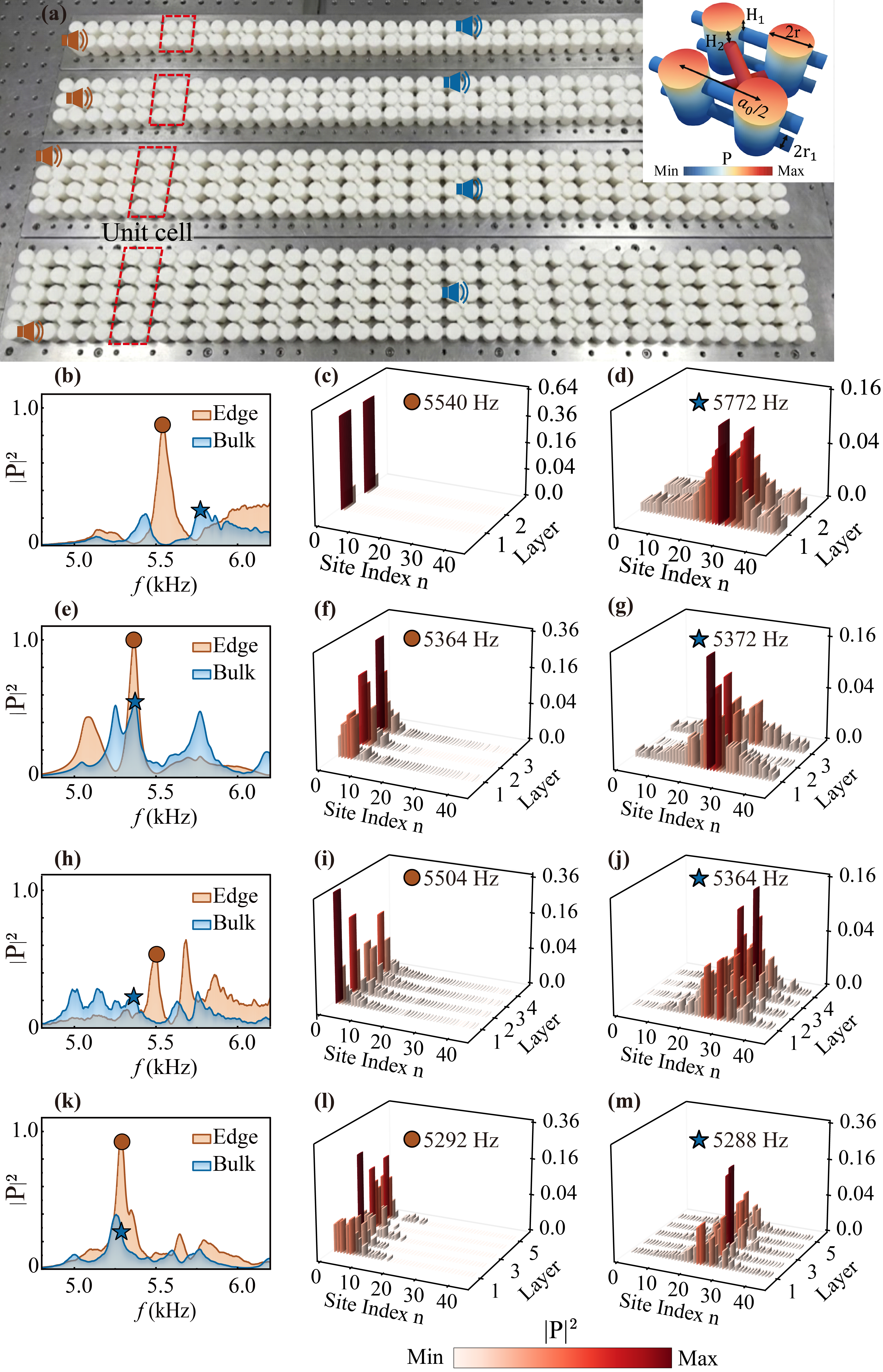}
\caption{Spectral and spatial characterization of multilayer acoustic lattices.
(a) Photograph of the fabricated experimental samples. The inset provides a magnified view of the resonator geometry, highlighting the engineered intra- and inter-layer coupling mechanisms. Colored speaker icons denote the selective excitation positions for edge (orange) and bulk (blue) states.
(b, e, h, k) Measured acoustic DOS for stack depths of $N = 2, 3, 4, \text{and } 5$ layers, respectively. The orange and blue shaded regions delineate the spectral domains dominated by edge-localized modes and bulk-distributed modes.
(c, f, i, l) Measured spatial acoustic pressure-field profiles $|P|^2$ for representative topological states (marked by orange circles in the spectra), demonstrating robust localization at the lattice boundaries.
(d, g, j, m) Spatial pressure-field profiles of the corresponding bulk states (marked by blue stars), exhibiting delocalized distributions across the primary lattice sites.}
\label{fig3}  
\end{figure}

\textit{Observation of Topological Boundary States and BICs in Acoustic Lattices}---We experimentally probe our theoretical predictions by fabricating the corresponding acoustic lattice structures using 3D printing. The fabricated lattices are shown in Fig.~\ref{fig3} (a), with geometric parameters: lattice constant $a_0 = 6\ \text{cm}$, height $H = 3\ \text{cm}$, $H_1=0.2H$, $H_2=0.3H$, main cavity radius $r = 0.9\ \text{cm}$, and connecting cavity radius $r_1 = 0.3\ \text{cm}$. The inter-cavity coupling strength is tuned via the vertical offset of the connecting cavities from the main cavity mid-height: $H/2 - H_1$ for intra-layer coupling and $H/2 - H_2$ for inter-layer coupling (see Fig.~\ref{fig3}(a)). This interlayer crossing coupling not only enlarges the unit cell of the underlying one-dimensional metallic lattice but, more importantly, enforces chiral symmetry in the composite system. Consequently, the acoustic lattices capture the essential topological features predicted by the tight-binding model. 

{\blue{To experimentally map these states, acoustic waves are injected using a localized single-point source (a loudspeaker). As denoted by the speaker icons in Fig. 4(a), this source is strategically positioned at either the lattice boundary or the lattice center to selectively excite the topological edge states or the bulk continuum modes, respectively. The resulting steady-state spatial acoustic pressure-field profiles are then mapped by systematically scanning a probe microphone through the main cavities across the entire structure.}}

The density of states (DOS) in Figs.~\ref{fig3}(b) and (h) reveals in-gap edge states for the 2- and 4-layer acoustic lattices. Their pressure-field distributions---shown for the edge states in Figs.~\ref{fig3}(c) and (i) and for the bulk states in Figs.~\ref{fig3}(d) and (j)---agree with the tight-binding model predictions. For the 3- and 5-layer lattices, the systems reside in a topological gapless phase with protected edge states, as previously indicated by the entanglement spectrum and open-boundary tight-binding calculations. Experimentally, we observe these protected edge states in Figs.~\ref{fig3}(f) and (l), with the corresponding bulk states shown in Figs.~\ref{fig3}(g) and (m). The measured density of states in Figs.~\ref{fig3}(e) and (k) confirm that the spectral weight of these edge states is embedded within the bulk continuum, confirming their nature as topological BICs.

In summary, our acoustic measurements directly confirm the existence of both gapped topological edge states (for $N=2,4$) and gapless topological BICs (for $N=3,5$), validating the theoretical predictions. The agreement between experiment and theory demonstrates that symmetry engineering via interlayer coupling is an effective strategy for inducing topology in stacked trivial systems.

\textit{Conclusion and Outlook}---In conclusion, we have demonstrated that symmetry-guided stacking of trivial building blocks can generate rich topological phases, including gapped systems with protected edge states and gapless systems hosting topological BICs. Our approach relies on engineered interlayer couplings that enforce both chiral and interlayer reflection symmetries, with the layer number $N$ serving as a discrete tuning parameter. Crucially, the reflection symmetry ensures the protection of BICs in odd-layer systems by decoupling localized boundary states from the bulk continuum via parity orthogonality. These topological states are confirmed by numerical simulations, symmetry analysis, and experimental observations in 3D-printed acoustic lattices.

These results establish a versatile framework for creating topological matter from trivial components, without requiring intrinsic material topology or complex moir\'e engineering. By leveraging the interplay between sublattice and spatial symmetries, we provide a robust pathway for capturing bound states within the radiation continuum. Given the large design space of classical wave systems and the diversity of symmetry-protected topological classes, our method paves the way for discovering and engineering new topological phases. Future directions include extending the approach to higher dimensions, incorporating non-Hermitian effects, and implementing these design principles in photonic, mechanical, and electronic metamaterials for robust wave control and device applications.

\textit{Acknowledgments}---This work was supported by the National Natural Science Foundation of China (Grants No.~92263208 and No.~12304494), the National Key R\&D Program of China (Grants No.~2022YFA1404400 and No.~2022YFA1404403), the Shanghai Science and Technology Committee (Grant No.~21JC1405600), the Research Grants Council of Hong Kong SAR (Grant No.~AoE/P-502/20), and the Fundamental Research Funds for the Central Universities.

\begin{quote}
$^{\chi}$KaiChenPhys@tongji.edu.cn, 
$^{*}$zhmgu@tongji.edu.cn, 
$^{\dagger}$jiezhu@tongji.edu.cn
\end{quote}

 \bibliographystyle{apsrev4-2}
\bibliography{nonab}
 \onecolumngrid

\section*{Supplemental Material for Layer-number parity induced topological phase transition}
\appendix 
\section{Entanglement Spectrum and Eigenvalues of the Correlation Matrix}
In this section, we provide a brief derivation of why the eigenvalues of the correlation matrix correspond to the entanglement spectrum for noninteracting systems; comprehensive derivations can be found in several references \cite{peschel2009reduced,chang2020entanglement}. The correlation matrix is defined as

\begin{equation}
C_{\alpha\beta} \equiv \operatorname{Tr}(\rho_A \phi_{\alpha}^\dagger \phi_{\beta}) = \langle G | \phi_{\alpha}^\dagger \phi_\beta | G \rangle,
\label{eq:s1}
\end{equation}
where $| G \rangle$ denotes the ground state, and $\alpha$, $\beta$ label the positions of lattice sites. The ground state can be expressed as
\begin{equation}
| G \rangle = \prod_{n \in \text{occ}} \left( \sum_{\alpha} \psi_{n,\alpha} \phi_{\alpha} \right)^\dagger | 0 \rangle,
\label{eq:s2}
\end{equation}
where $| 0 \rangle$ is the vacuum state, and $H | \psi_n \rangle = \epsilon_n | \psi_n \rangle$ with $| \psi_n \rangle$ denoting the $n$th eigenstate of the Hamiltonian $H$. Using this expression for the ground state together with the commutation relation $\{\phi_{\alpha}^\dagger , \phi_{\beta}\}=\delta_{\alpha\beta}$ and the condition $\phi_{\alpha}| 0\rangle = 0$, the correlation matrix simplifies to
\begin{equation}
C_{\alpha\beta} = \sum_{n \in \text{occ}} \psi_{n,\alpha} \psi_{n,\beta}^*,
\label{eq:s3}
\end{equation}
where $\psi_{n,\beta}^*$ denotes the complex conjugate of $\psi_{n,\beta}$.

The correlation matrix $C$ and the entanglement Hamiltonian $H^E$ appearing in the reduced density matrix $\rho_A \propto \exp\left(-\sum_{\alpha\beta} \phi_{\alpha}^\dagger H_{\alpha\beta}^{E} \phi_\beta\right),$ can be simultaneously diagonalized, leading to the relation
\begin{equation}
C = \frac{e^{-H^E}}{1 + e^{-H^E}}.
\label{eq:s4}
\end{equation}
Consequently, the entanglement entropy for subsystem $A$ is given by
\begin{equation}
S_A = - \sum_{\delta} \left[ \varepsilon_\delta \ln \varepsilon_\delta + (1 - \varepsilon_\delta) \ln (1 - \varepsilon_\delta) \right],
\label{eq:s5}
\end{equation}
where $\varepsilon_\delta$ are the eigenvalues of $C$. Hence, the eigenvalues of the correlation matrix directly encode the entanglement spectrum.
\label{app1}

\section{Gapless nature for odd number of layers}

In this section, we demonstrate that the model considered in the main text becomes \textbf{gapless} when the number of layers $N$ is odd.

Consider a stack of $N$ layers of a trivial subsystem, each possessing chiral symmetry. The momentum-space Hamiltonian takes the block-tridiagonal form
\begin{equation}
H(k) =
\begin{pmatrix}
H_{\mathrm{onsite}}(k) & T      & 0      & \cdots & 0      \\
T^{\dagger}            & H_{\mathrm{onsite}}(k) & T      & \cdots & 0      \\
0                      & T^{\dagger} & H_{\mathrm{onsite}}(k) & \ddots & \vdots \\
\vdots                 & \vdots & \ddots & \ddots & T      \\
0                      & 0      & \cdots & T^{\dagger} & H_{\mathrm{onsite}}(k)
\end{pmatrix},
\label{eq:Hk_full}
\end{equation}
where the intralayer Hamiltonian is  
\begin{equation}
H_{\mathrm{onsite}}(k) = t\,(1+\cos k)\,\sigma_x + t\sin k\,\sigma_y,
\label{eq:H_onsite}
\end{equation}
and the interlayer hopping matrix is  
\begin{equation}
T \equiv t_1 \sigma_x.
\label{eq:T_matrix}
\end{equation}

At the time-reversal invariant momenta $k = \pm\pi$, we have $H_{\mathrm{onsite}}(\pm\pi)=0$, and the Hamiltonian reduces to  
\begin{equation}
H(\pm\pi) =
\begin{pmatrix}
0 & T      & 0      & \cdots & 0      \\
T^{\dagger} & 0      & T      & \cdots & 0      \\
0 & T^{\dagger} & 0      & \ddots & \vdots \\
\vdots & \vdots & \ddots & \ddots & T      \\
0 & 0      & \cdots & T^{\dagger} & 0
\end{pmatrix}.
\label{eq:H_pi}
\end{equation}

Direct inspection shows that $H(\pm\pi)$ possesses zero-energy eigenstates of the form  
\begin{equation}
\Psi(\pm\pi) = \bigl[ \psi_1,\;0,\;\psi_3,\;0,\;\dots,\;0 \bigr]^{\!\mathsf{T}},
\qquad \psi_1 = -\psi_3,
\label{eq:zero_mode}
\end{equation}
where the superscript $\mathsf{T}$ denotes the transpose.

Because the system respects chiral symmetry, the zero modes at $\pm\pi$ must appear in even numbers. Combined with the continuity of the band structure in $k$-space, this implies that for \textbf{odd} $N$, the bulk spectrum remains gapless. 

Together with the entanglement-spectrum analysis presented in the main text, this result confirms the existence of topological bound states in the continuum (BICs) in odd-layer systems.

{\blue{
\section{Entanglement Entropy of the 3-Layer and 5-Layer Gapless Systems}
\label{app2}
\begin{figure}[h]
\renewcommand{\thefigure}{S1} 
\includegraphics[width=1\columnwidth]{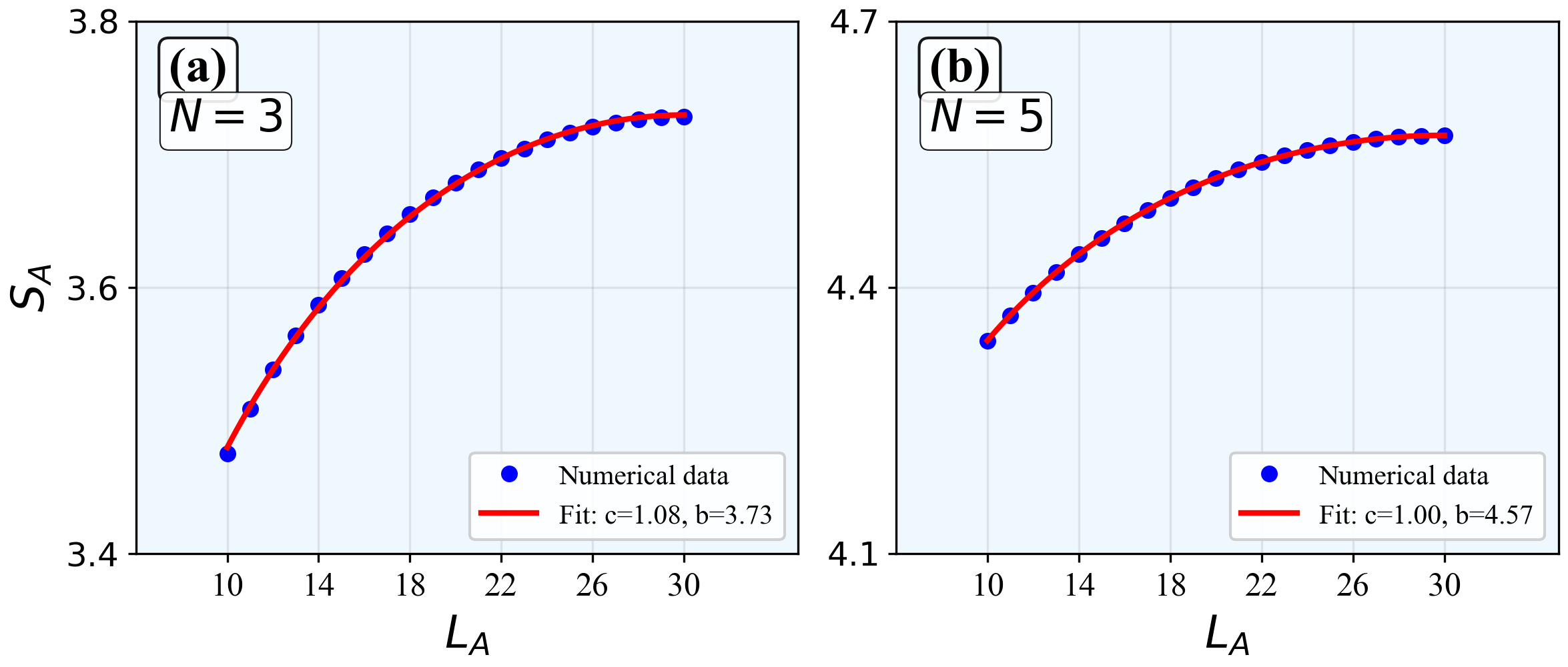}
\caption{(a, b) Entanglement entropy as a function of subsystem size for the 3-layer and 5-layer systems, respectively. The parameters are $t=1$ and $t_1=1.34$.}
\label{ees}
\end{figure}
As described in the main text, the 3-layer and 5-layer systems are gapless and host topological BICs. It is well established by conformal field theory (CFT) that the entanglement entropy of a 1D gapless critical system scales logarithmically with the subsystem size $L_A$. Specifically, it follows the relation
\begin{equation}
S_A (L_A) = \frac{c}{3} \ln\left[ \sin\left(\frac{\pi L_A}{L}\right) \right] + b,
\end{equation}
where $c$ denotes the central charge, $L$ is the size of the full system, and $b$ is a non-universal constant. In this section, we calculate the entanglement entropy of our lattices to extract the central charge. As shown in Fig.~\ref{ees}, the numerically obtained entanglement entropy agrees exceptionally well with the CFT prediction, yielding a central charge of $c\approx1$.

\label{appEE}

\section{BICs Protected by Interlayer Reflection Symmetry}
In the previous section, we showed that systems with an odd number of layers are gapless; in this section, we show that interlayer reflection symmetry acts as a protection mechanism for the BICs in these systems. Since it is straightforward to generalize the proof to other cases, we use the three-layer case as a concrete example to demonstrate how interlayer reflection symmetry protects the BIC.

The system is constructed by stacking identical trivial systems with uniform nearest-neighbor interlayer coupling. Therefore, by treating each layer as a degree of freedom, the effective Hamiltonian can be written as:

\begin{equation}
H_{eff} = I_{3\times 3} \otimes H_0 + M\otimes T,
\end{equation}

where $H_0 = I_{N_u \times N_u} \otimes \sigma_x t + \left[ I_\Delta \otimes \frac{t}{2} (\sigma_x + i \sigma_y) + \text{h.c.} \right]$, and the elements of $I_\Delta$ are defined as $(I_\Delta)_{mn} = \delta_{m+1,n}$. The matrix $T = I_{N_u \times N_u} \otimes \sigma_x t_1$, with $N_u$ denoting the number of unit cells in each layer, and the matrix $M$ is given by
\begin{equation}
M =
\begin{pmatrix}
0 & 1      & 0        \\
1 & 0      & 1      \\

0 & 1      &  0
\end{pmatrix}.
\end{equation}

The transverse layer coupling is governed by the matrix $M$, which yields an antisymmetric eigenstate $\psi_{0} = [1, 0, -1]^T$ with eigenvalue $\lambda=0$, and symmetric eigenstates $\psi_{\pm} = [1, \pm\sqrt{2}, 1]^T$ with eigenvalues $\lambda=\pm\sqrt{2}$. The cross-coupled system can be mapped to decoupled effective 1D chains with intercell coupling $t$ and an effective intracell coupling $t_{\mathrm{eff}} = t + \lambda t_1$. 

The existence of a gapless propagating bulk mode at exactly zero energy requires the critical condition $t_{\mathrm{eff}} = t$, which strictly selects the antisymmetric channel ($\lambda=0$). Conversely, the symmetric channel characterized by $\lambda=-\sqrt{2}$ yields an effective hopping $t_{\mathrm{eff}} = t - \sqrt{2}t_1$. Provided that $|t - \sqrt{2}t_1| < |t|$ (in the main text, we set $t=1$ and $t_1=1.34$), this specific chain is driven into the topological phase, naturally hosting highly localized zero-energy boundary states, which constitute the BIC. In the five-layer system, the bulk zero mode corresponds to the symmetric channel ($\lambda=0$), while the BIC corresponds to the antisymmetric channel ($\lambda=-1$).

Consequently, the zero-energy bulk continuum and the zero-energy topological boundary state inherit orthogonal spatial parities under interlayer reflection. Due to this symmetry mismatch, their interaction matrix element strictly vanishes. Hence, the topological mode is dynamically decoupled from the radiative continuum in the absence of symmetry-breaking perturbations.

\label{app3}

\section{Transmission Spectra via the Non-Equilibrium Green's Function}

To further confirm the conclusions established in the preceding section, we calculate the transmission spectra using the Non-Equilibrium Green's Function method \cite{datta1997electronic}. This method allows us to extract the scattering properties of the finite lattice by coupling it to semi-infinite propagating leads. The retarded Green's function governing wave propagation through the system is given by
\begin{equation}
G(E)= [E\mathbf{I} - H_{s}-\Sigma_L (E)-\Sigma_R (E)]^{-1},
\end{equation}
where $H_s$ denotes the Hamiltonian of the system of interest, and $\mathbf{I}$ is the identity matrix. The terms $\Sigma_{L/R}(E)$ are the left and right self-energies, which rigorously capture the coupling between the finite system and the radiative environment. These self-energies are expressed as
\begin{equation}
\Sigma_{L/R} (E) = g_{\mathrm{surface}}(E) \left( \tau_{L/R} \tau_{L/R}^\dagger \right),
\end{equation}
where the surface Green's function of the uniform 1D leads is $g_{\mathrm{surface}}(E) = \frac{E - \sqrt{E^2 - 4\alpha^2}}{2\alpha^2}$, with $\alpha$ denoting the lead-system coupling amplitude (set to $\alpha=t$). The column vector $\tau_{L/R}$ defines the spatial injection profile across the transverse layers, dictating exactly which synthetic modes the incident wave will excite.

The total transmission probability $T(E)$ is obtained via:
\begin{equation}
T(E) = 4 \mathrm{Tr} \left[ \mathrm{Im}[\Sigma_{L}(E)] \, G(E) \, \mathrm{Im}[\Sigma_{R}(E)] \, G^\dagger(E) \right].
\end{equation}

To systematically demonstrate the dynamic decoupling of the bound states, we engineer the injection vector $\tau$ to probe different symmetries of the lattice. For the 3-layer system, we employ two distinct incident wave configurations:
\begin{equation}
\tau_{\mathrm{naive}} = \begin{pmatrix} 1 \\ 0 \\ 0 \end{pmatrix}, \quad \quad \tau_{\mathrm{anti}} = \frac{1}{\sqrt{2}} \begin{pmatrix} 1 \\ 0 \\ -1 \end{pmatrix}.
\end{equation}
The single-layer injection ($\tau_{\mathrm{naive}}$) breaks the transverse layer symmetry, containing both symmetric and antisymmetric components. Consequently, it artificially couples to the localized BIC, triggering a Fano resonance and yielding a transmission dip at exactly $E=0$. Conversely, because the 3-layer BIC is rigorously symmetric under interlayer reflection, the coherent antisymmetric injection ($\tau_{\mathrm{anti}}$) is mathematically orthogonal to it. This wave exclusively excites the gapless bulk continuum, bypassing the BIC entirely and resulting in perfect reflectionless transmission ($T=1$) at zero energy.

\begin{figure}[h]
\renewcommand{\thefigure}{S2} 
\includegraphics[width=1\columnwidth]{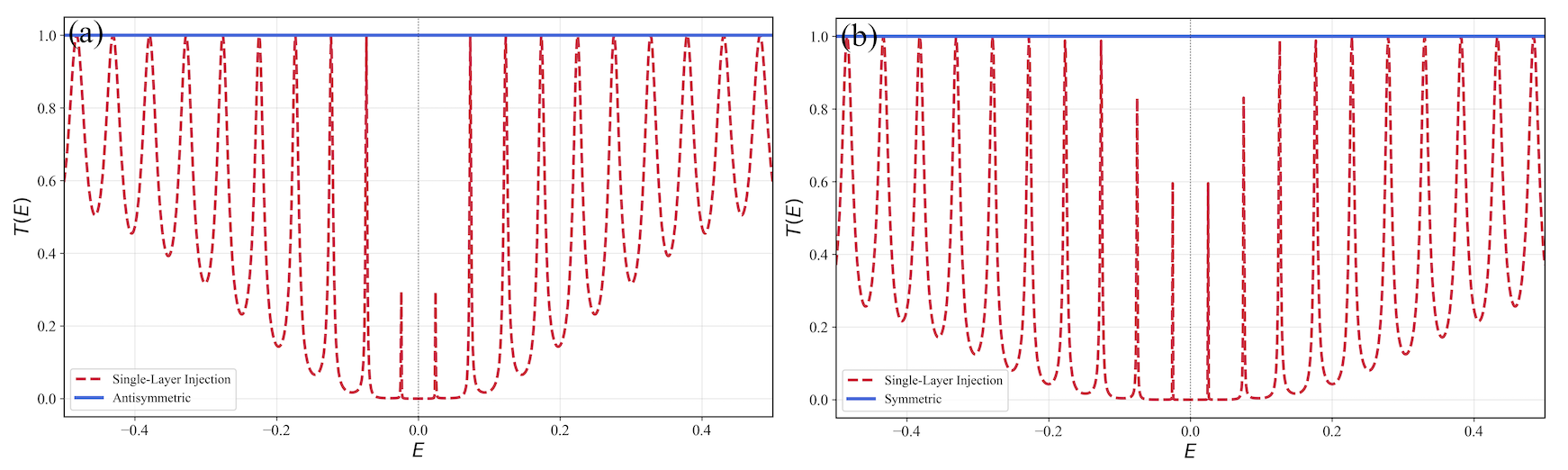}
\caption{(a, b) Transmission spectra for the 3-layer and 5-layer systems, respectively.}
\label{figs2}
\end{figure}

Crucially, in the 5-layer system under our parameter regime, the topological channel hosts an antisymmetric BIC ($\lambda=-1$, transverse profile $[1, -1, 0, 1, -1]^T$), whereas the gapless continuum ($\lambda=0$) exhibits a symmetric profile. Because the spatial parities are inverted compared to the 3-layer case, the incident wave must be adapted to selectively excite this symmetric continuum channel. The corresponding 5-layer coherent injection vector is:
\begin{equation}
\tau_{\mathrm{sym}} = \frac{1}{\sqrt{3}} \begin{pmatrix} 1 \\ 0 \\ -1 \\ 0 \\ 1 \end{pmatrix}.
\end{equation}
By matching the exact node structure of the $\lambda=0$ continuum, $\tau_{\mathrm{sym}}$ ensures the incident wave remains strictly orthogonal to the topological boundary mode. We numerically calculate the transmission spectra for these configurations (shown in Figure ~\ref{figs2}), which confirm the complete suppression of Fano scattering and demonstrate perfect dynamic decoupling in accordance with our theoretical parity analysis.

\label{app4}
}}
\section{Topological in-gap states and BICs under alternative parameters}

\begin{figure}[h]
\renewcommand{\thefigure}{S3} 
\includegraphics[width=1\columnwidth]{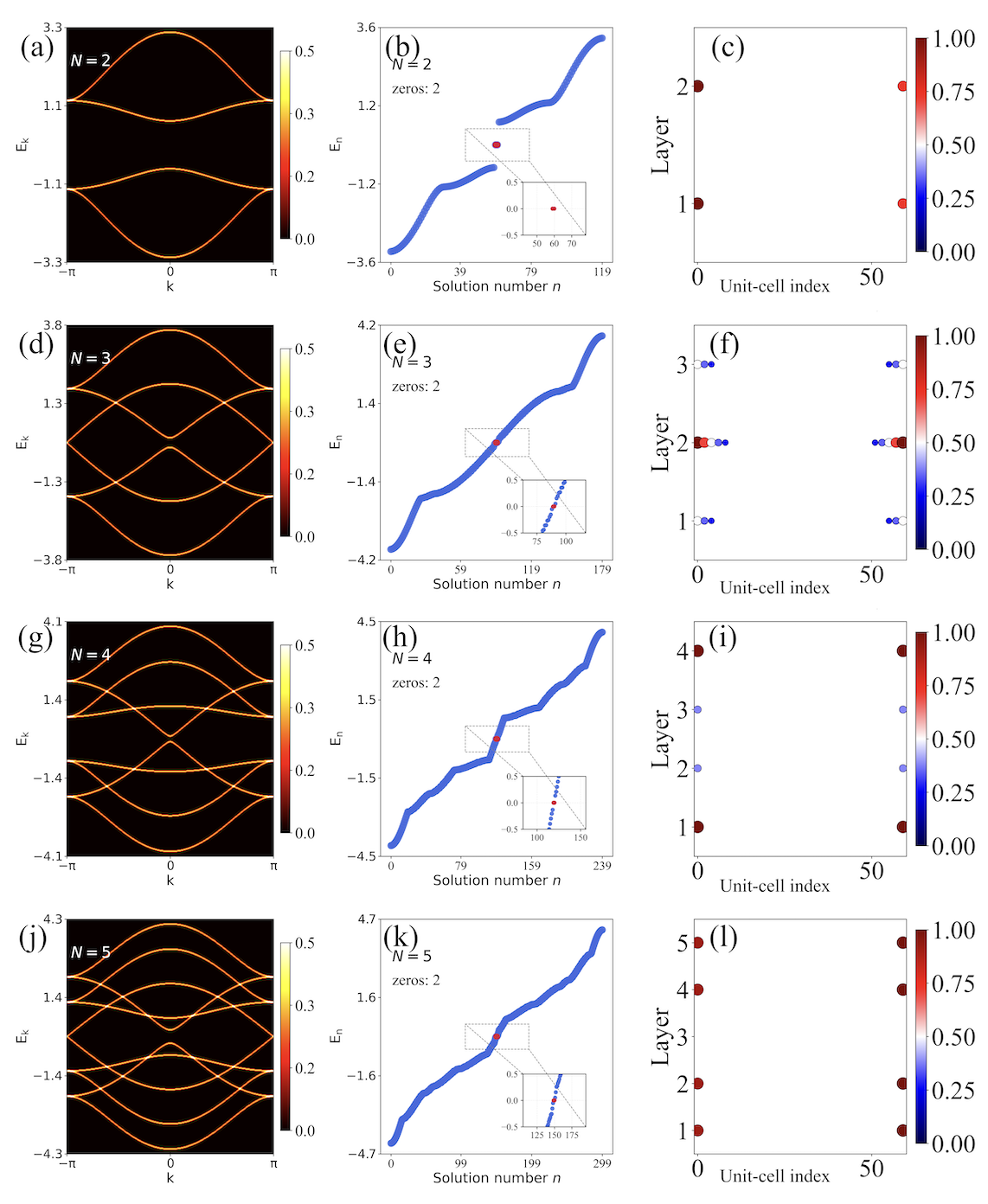}
\caption{(Color online) Transition between topological in-gap states and BICs in multi-layer systems.
(a--c) Bulk band structures weighted by the spectral function $\text{A(k, E)}$, energy spectra for finite systems under open boundary conditions, and spatial distribution of the magnitude of characteristic wavefunctions $|\psi|$ for 2-layer configurations. 
(d--f), (g--i), and (j--l) show corresponding results for 3, 4, and 5-layer configurations, respectively. 
The topological states (red dots) appear as in-gap edge states for even-layer systems ($N=2, 4$) and as BICs embedded within the continuum for odd-layer systems ($N=3, 5$). 
Parameters: $t = 1$, $t_1 = 1.3$.}
\label{figs1.3}
\end{figure}

{\blue{In this section, we demonstrate the layer-parity-selected topological in-gap states and BICs using different parameters from the main text to show that the effect is not fine-tuned. As shown in Figure ~\ref{figs1.3}, the topological in-gap states exist in the 2-layer and 4-layer systems, and the BICs occur in the 3-layer and 5-layer systems.}}

\label{app6}

\section{Comsol-simulated eigenspectra and eigenmodes}
\begin{figure}[h]
\renewcommand{\thefigure}{S4} 
\includegraphics[width=1\columnwidth]{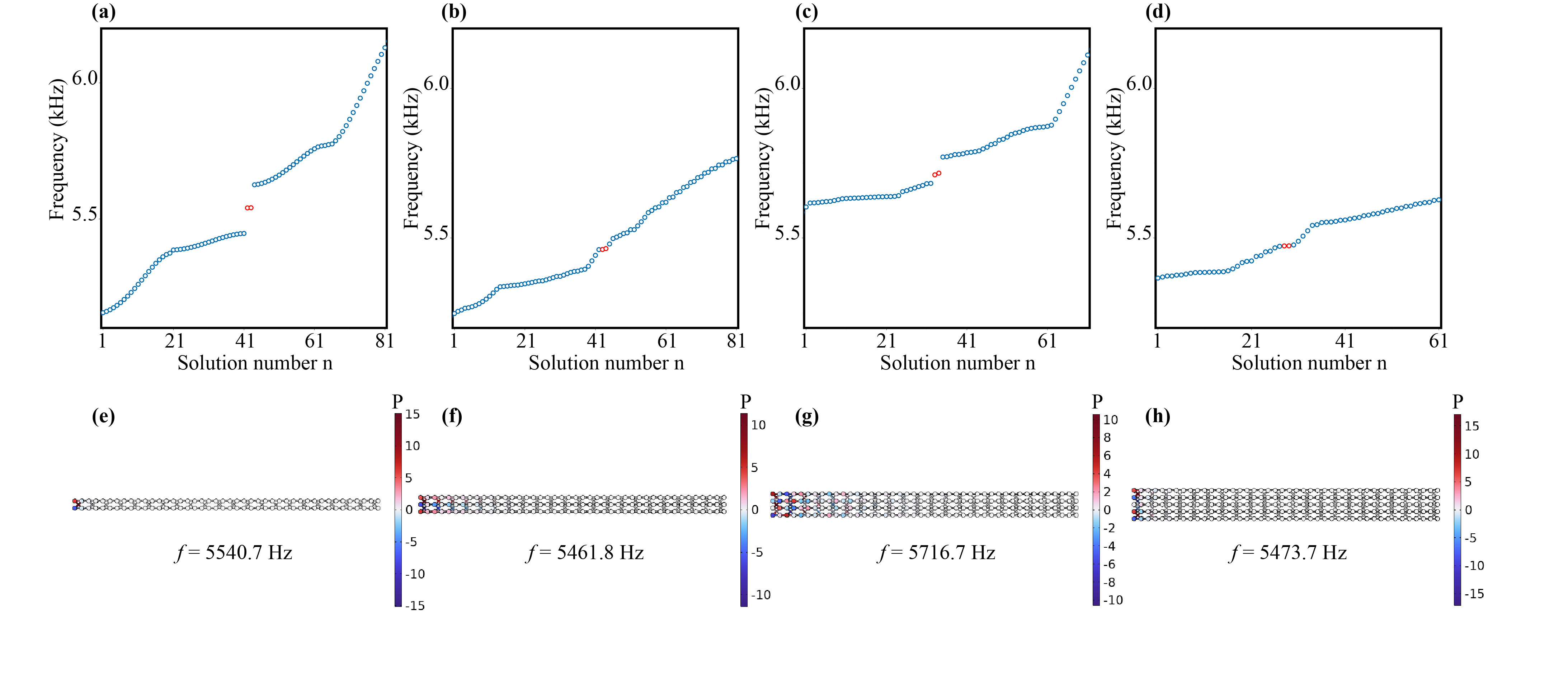}
\caption{(a, c) Eigenspectra for the 2- and 4-layer structures; (e, g) show the corresponding in-gap edge states, which are highlighted in red in (a) and (c). (b, d) Eigenspectra for the 3- and 5-layer structures; (f, h) show the corresponding BICs, which are highlighted in red in (b) and (d).}
\label{figs1}
\end{figure}

In this section, we present the eigenspectra and eigenmodes obtained from Comsol simulations. As shown in Figure ~\ref{figs1}, the 2- and 4-layer systems host in-gap edge states, while the 3- and 5-layer systems exhibit BICs. These observations agree with the predictions based on the tight-binding model in the main text.

\label{app5}

\end{document}